\documentclass[sigplan,10pt]{acmart}
\AtBeginDocument{%
  }

\copyrightyear{2026}
\acmYear{2026}
\setcopyright{cc}
\setcctype{by}
\acmConference[EUROSYS '26]{21st European Conference on Computer Systems}{April 27--30, 2026}{Edinburgh, Scotland Uk}
\acmBooktitle{21st European Conference on Computer Systems (EUROSYS '26), April 27--30, 2026, Edinburgh, Scotland Uk}
\acmPrice{}
\acmDOI{10.1145/3767295.3769353}
\acmISBN{979-8-4007-2212-7/26/04}

\usepackage{graphicx}
\usepackage{dcolumn}
\usepackage{bm}
\usepackage{times}
\usepackage{outlines}
\usepackage{breakurl}
\usepackage{amsthm}
\usepackage{amsfonts}
\usepackage{hyperref}
\usepackage{ulem}
\usepackage{xspace}
\usepackage{paralist}
\usepackage[dvipsnames]{xcolor}
\usepackage{enumitem}
\usepackage{outlines}
\usepackage{subcaption}
\usepackage{booktabs}
\usepackage{fnpct}
\usepackage{xparse}
\usepackage{xstring}

\ExplSyntaxOn
\NewDocumentCommand{\btexttt}{m}
 {
  \texttt{
    \tl_set:Nn \l_tmpa_tl {#1}
    \tl_map_function:NN \l_tmpa_tl \__insert_breakpoints:n
  }
 }

\cs_new:Nn \__insert_breakpoints:n
 {
  \str_if_in:nnTF {./-} {#1}
   {#1\hspace{0pt}}
   {#1}
 }
\ExplSyntaxOff

\newcommand{\hide}[1] 
{
\ifthenelse{\boolean{false}}{#1}{}
}
\newcommand{\secref}[1]{\S\ref{#1}}
\newtheoremstyle{mystyle} 
  {.3\topsep}               
  {.3\topsep}               
  {\itshape}              
  {}                      
  {\bfseries}             
  {.}                     
  {.5em}                  
  {}                      

\theoremstyle{mystyle}

\newcommand{\eat}[1]{}
\usepackage{algorithm, algorithmicx}
\usepackage[noend]{algpseudocode}
\usepackage{multirow}
\usepackage{subcaption}
\usepackage{pifont}
\usepackage{enumitem}
\usepackage{refcount}
\setlist[itemize]{itemsep=0pt, partopsep=0pt, parsep=1pt, topsep=1pt}
\setlist[enumerate]{itemsep=0pt, partopsep=0pt, parsep=1pt, topsep=1pt}

\usepackage{minted}
\setminted{
  python3=true,
  fontsize=\footnotesize,
  autogobble,
  xleftmargin=1em,
  numbersep=4pt,
  highlightcolor=lightgray,
}
\setmintedinline{fontsize=\small}
\newmintedfile[pythoncode]{python}{}
\newmintedfile[yamlcode]{yaml}{
    baselinestretch=1.2
}

\usepackage{booktabs} 
\usepackage{siunitx} 
\newcommand{\sys}{\btexttt{SkyWalker}\xspace}

\newcommand{\MyPara}[1]{\vspace{1mm}\textbf{\textit{#1}}~}

\begin{document}

\title{SkyWalker: A Locality-Aware Cross-Region Load Balancer for LLM Inference}

\author{
Tian Xia$^{\dagger}$\;
Ziming Mao$^{\dagger}$\;
Jamison Kerney$^{\dagger}$\;
Ethan J. Jackson$^{\dagger}$\;
Zhifei Li$^{\S}$\;
\\
Jiarong Xing$^{\dagger\P}$\;
Scott Shenker$^{\dagger\diamond}$\;
Ion Stoica$^{\dagger}$\; 
\\
\emph{$^\dagger$UC Berkeley}
\emph{$^\S$Renmin University of China}
\emph{$^\P$Rice University}
\emph{$^\diamond$ICSI}
}

\renewcommand{\shortauthors}{Xia et al.}
\renewcommand{\shorttitle}{SkyWalker: A Locality-Aware Cross-Region Load Balancer for LLM Inference}
\renewcommand{\authors}{Tian Xia, Ziming Mao, Jamison Kerney, Ethan J. Jackson, Zhifei Li, Jiarong Xing, Scott Shenker, Ion Stoica}

\begin{abstract}
Serving Large Language Models (LLMs) efficiently in multi-region setups remains a challenge.
Due to cost and GPU availability concerns, providers typically deploy LLMs in multiple regions using instance with long-term commitments, like reserved instances or on-premise clusters, which are often underutilized due to their region-local traffic handling and diurnal traffic variance.
In this paper, we introduce \sys, a multi-region load balancer for LLM inference that aggregates regional diurnal patterns through cross-region traffic handling. By doing so, \sys enables providers to reserve instances based on expected \emph{global} demand, rather than peak demand in each individual region. Meanwhile, \sys preserves KV-Cache locality and load balancing, ensuring cost efficiency without sacrificing performance.
\sys achieves this with a cache-aware cross-region traffic handler and a selective pushing based load balancing mechanism.
Our evaluation on real-world workloads shows that it achieves $1.12$-$2.06\times$ higher throughput and $1.74$-$6.30\times$ lower latency compared to existing load balancers, while reducing total serving cost by $25\%$.
\end{abstract}

\ccsdesc[500]{Computing methodologies~Distributed artificial intelligence}
\ccsdesc[500]{Computing methodologies~Distributed computing methodologies}

\keywords{Load Balancing, AI Serving, Multi-Region, Cloud Computing}

\maketitle

\section{Introduction}

Large Language Models (LLMs) have seen rapid growth in usage in recent years. With increasingly advanced capabilities, they have been adopted across a wide range of domains, including virtual assistants~\cite{appleChatGPTiPhone, googleGeminiAssistant}, code generation~\cite{githubCopilotFeatures, awsCodeWhispererGA, tabnine, replitGhostwriterIntro, windsurfAI,cursor}, and information search~\cite{consensus2025,pplx-ai}. These applications now serve billions of users worldwide~\cite{openai-800m-user}. 

\begin{figure}[ht]
  \centering
  \includegraphics[width=\linewidth]{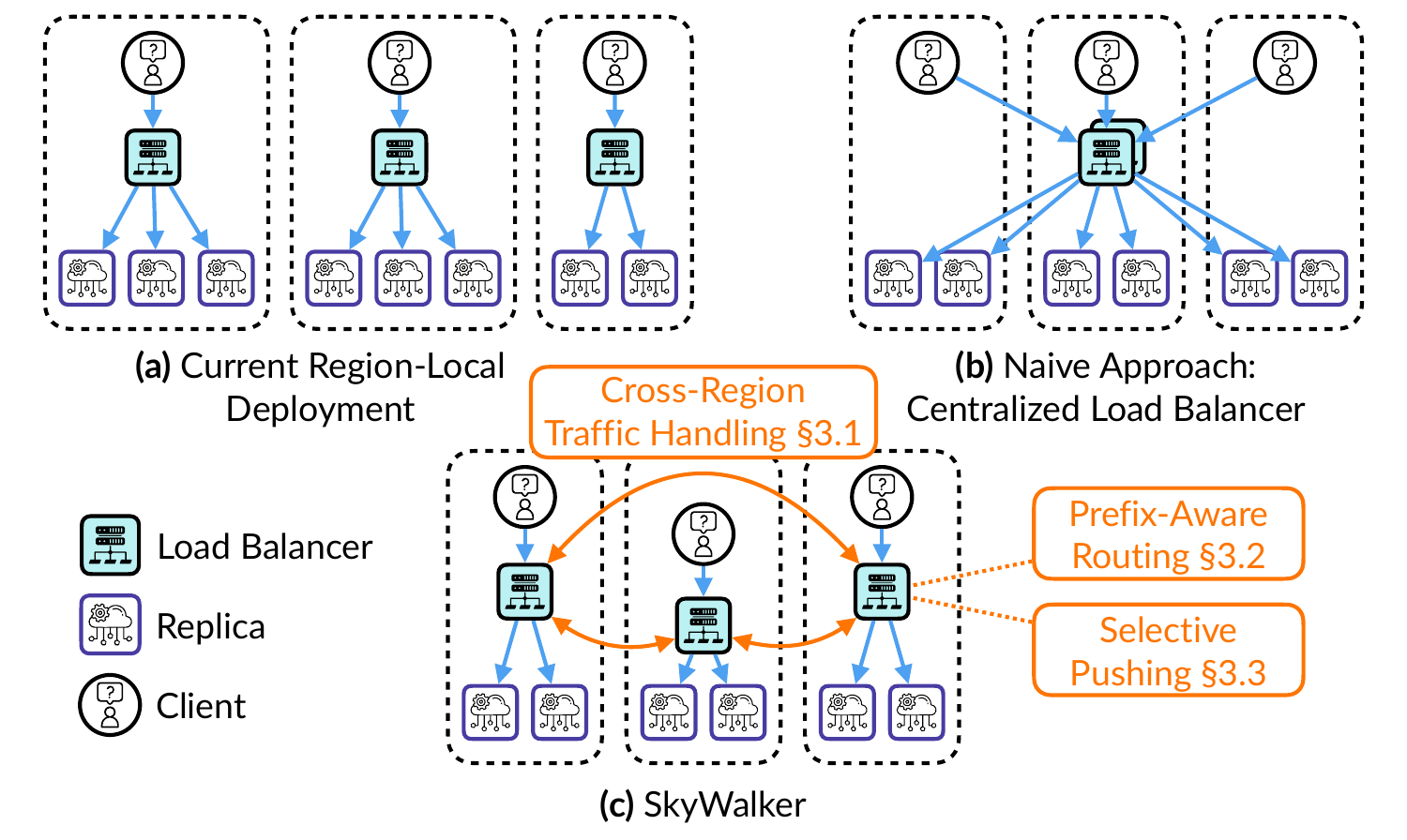}
  \caption{(a) Current region-local deployment is cost inefficient due to provisioning each region for its peak load. 
  (b) A naive centralized load balancer can reduce costs, but it introduces high cross-region latency, performance bottlenecks, and single point of failure. (c) \sys reduce multi-region serving costs by enabling cross-region traffic handling without sacrificing performance.}
  \label{fig:overview}
\end{figure}

Serving LLMs at this scale requires operators to manage GPU scarcity while optimizing cost, latency, and throughput. To meet these demands, LLM providers often deploy infrastructure across multiple geographical regions. The most common practice today is to deploy GPU instances with long-term commitments in each region, like reserved instances~\cite{googleComputeReservations} or on-premise clusters~\cite{sherwood2024gpuhoarding,shilov2024bytedance,xai-gpu-cluster}, as shown in Figure~\ref{fig:overview}(a). The long-term commitments reduce GPU costs compared to purchasing on-demand instances with autoscaling (\S\ref{ssec:muti-region-serve}) and ensure GPU availability at any time, while the region-local deployment improves service quality by placing compute resources closer to end users.

\begin{figure}[t]
  \centering
  \includegraphics[scale=0.5]{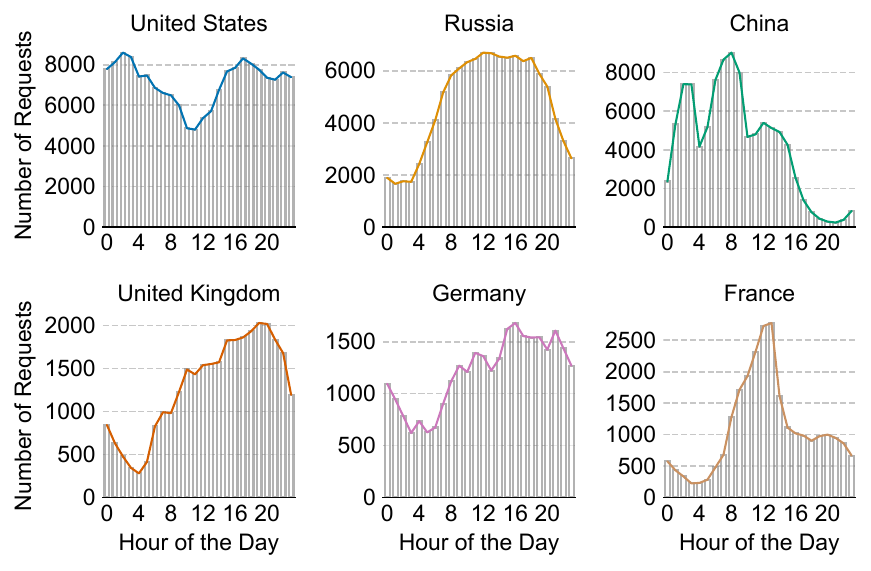}
  \caption{Regional traffic demands shift over time in multi-region LLM serving. Data is from the WildChat~\cite{zhao2024wildchat} trace.}
  \label{fig:diurnal-pattern}
\end{figure}

However, this approach introduces significant provisioning and cost management challenges. These reserved instances or on-premise clusters are inflexible---providers cannot increase or decrease capacity within a region as demand shifts. As a result, providers must allocate enough instances in each region to meet peak demand, which can result in high and often wasted costs. This inefficiency arises from shifting regional traffic patterns that follow distinct daily diurnal cycles: each region experiences peak inference load at certain times of day, with lower demand during off-peak hours (Figure~\ref{fig:diurnal-pattern}). The challenge is further exacerbated in multi-region deployments, where providers must provision for peak load in every region independently. This leads to resource fragmentation and underutilization during off-peak hours.

Ideally, providers would perfectly meet capacity with demand while paying the lower costs of reserved instances or on-premise clusters. However, achieving this ideal is difficult when regional provisioning is viewed in isolation. Providers must either provision for peak demand for all regions or pay for the flexibility of on-demand instances and take the risk of GPU unavailability.

In this paper, we suggest relaxing the regional rigidity of current approaches. Instead of attempting to match regional capacity to regional demand, providers make reservations for peak \textit{global} demand and partition those reservations across the regions closest to their users. Following this approach, when a region is overloaded, it can offload requests to other regions with excess capacity. By reserving for global peak demand and enabling cross-region traffic handling, a system can improve GPU utilization and reduce overall serving costs. Unfortunately, this cannot be achieved by simply deploying centralized load balancers in a single zone (Figure~\ref{fig:overview}(b), \cite{srivatsa2024prebleefficientdistributedprompt, sglang-router}), as this introduces high latency due to cross-region communication and creates both a performance bottleneck and a single point of failure.

It is important to note that cross-region traffic handling, as proposed in this work, has not been a common practice for traditional workloads such as webpage rendering or search engine queries. In these cases, the processing time per request is typically very short~\cite{vesper-nsdi}, often just several tens of milliseconds, and usually smaller than the cross-region network latency that is up to 200 milliseconds~\cite{mao2024skyserveservingaimodels}. As a result, requests are handled entirely within the region closest to the client. In contrast, LLM requests often require seconds, if not tens of seconds to complete~\cite{yu2025prism}. In this setting, cross-region latency represents only a small fraction of the total processing time. However, responsiveness remains important, and LLM providers still focus on optimizing Time-to-First-Token (TTFT) latency and aim to serve requests locally \emph{when capacity allows}.

To enable cross-region traffic handling without sacrificing performance, we design \sys, a cross-region load balancer for LLM inference. As shown in Figure~\ref{fig:overview}(c), \sys deploys at least one load balancer in each region as the first point of contact for requests, ensuring low-latency and avoiding centralized bottlenecks. These regional load balancers collaboratively coordinate traffic across regions to handle load imbalances.  Achieving this requires us to take into consideration two key challenges in multi-region load balancing: Key-Value Cache (KV Cache) awareness and LLM inference load unpredictability. We discuss both briefly below before expanding in the rest of the paper.

\MyPara{KV Cache awareness.} 
Modern LLM inference relies heavily on a KV Cache to reuse computation when requests share a common prefix. Previous work has demonstrated that routing such requests to the same GPU is critical for maximizing KV Cache utilization and achieving high throughput~\cite{zheng2024sglang,gim2024prompt, kwon2023efficientmemorymanagementlarge, srivatsa2024prebleefficientdistributedprompt}. However, these approaches have primarily assumed single load balancer deployments within a single zone. \sys overcomes the challenge of coordinating prefix-aware routing across multiple regions. It offers two routing algorithms to preserve KV Cache locality in a geo-distributed setting: a simple multi-region extension of consistent hashing based on user ID and session ID, and a more general multi-region prefix trie.

\MyPara{LLM inference load unpredictability.} 
We observe that the processing time and resource usage for LLM inference of a particular request is highly variable and unpredictable~\cite{zheng2023response, yu2025prism}. A single request can complete within seconds that requiring very little resources, or takes tens of seconds to complete while requiring several gigabytes of GPU memory. To make matters worse, due to the auto-regressive nature of LLM decoding, it is hard to predict which end of this spectrum a particular request will fall on~\cite{beigi2024rethinkinguncertaintycriticalreview}. As we show in section \secref{background-unknown-output-length}, this property of LLMs makes traditional load balancing policies such as least load first or round robin highly ineffective. To address this, \sys implements selective pushing based on pending requests, an algorithm we describe in section \secref{ssec:selective-pushing}, which improves performance by balancing load based on each replica's availability of admitting more requests in its continuous batch~\cite{orca280922}.

This paper proceeds as follows.  In section \S\ref{sec:background-motivation}, we describe the background of LLM serving, motivate the need for cross-region routing to lower cost, and identify new challenges in effectively doing so. In \S\ref{sec:skywalker-design}, we introduce the key design choices of \sys, including cross-region traffic handling, multi-region prefix-aware routing, and selective pushing to mitigate load imbalance. We evaluate \sys on three realworld workloads, and we observe that \sys achieves $1.12$-$2.06\times$ higher throughput and $1.74$-$6.30\times$ lower latency compared to existing load balancers. By using cross-region routing, we show that \sys is able to reduce $25\%$ of total cost compared to region-local deployment. In summary, this paper makes four contributions: 

\begin{itemize}[leftmargin=*, topsep=6pt]
    \item Identifying the need for cross-region traffic handling to aggregate diurnal patterns across multiple geographical regions and reduce global serving cost.
    \item Proposing two mechanisms to provide effective cross-region routing: prefix-aware routing to improve cache locality and performance, and selective pushing based on pending requests at each replica to reduce load imbalance.
    \item A comprehensive evaluation for \sys, compared to existing production and research systems across a variety of workloads.
    \item An open-source system \sys to stimulate further research on cross-region load balancing for LLMs.
\end{itemize}

\section{Background and Motivation}
\label{sec:background-motivation}

In this section, we first provide background on multi-region LLM serving. We then introduce the global cost reduction problem, which motivates cross-region load balancing. Finally, we discuss the key challenges in enabling cross-region load balancing without sacrificing performance.

\subsection{Background: Multi-Region LLM Serving}
\label{ssec:muti-region-serve}

\MyPara{LLM inference.}
LLM inference typically consists of two stages: prefill and decode. During the prefill stage, the model processes the initial input prompt and generates an internal KV Cache, which stores intermediate states necessary for subsequent token generation. Following prefill, the decode stage generates tokens auto-regressively, one token at a time, utilizing the KV Cache to accelerate inference. To improve GPU utilization and throughput, continuous batching~\cite{orca280922} is commonly employed, which dynamically groups incoming requests to reduce idle time. Due to strict latency SLOs, practitioners typically run online LLM serving on dedicated clusters to prevent interference from other jobs (\secref{related-work}).

\MyPara{Scaling LLM serving to multiple regions.}
To scale effectively, major providers~\cite{morikawa_2023} leverage GPU resources across multiple geographical regions to serve their users, a practice we refer to as \emph{multi-region serving} in this paper. Multi-region serving offers two key benefits: (1) improved latency by deploying LLM replicas closer to users; (2) reduces the risk of GPU shortages in any single region by diversifying GPU usage across multiple regions~\cite{mao2024skyserveservingaimodels, yang2023skypilot}.

\MyPara{Provisioning GPUs via long-term commitments.}
\label{long-term-commitment}
The substantial GPU demands of LLM serving, combined with ongoing GPU scarcity, have led providers to primarily deploy reserved or on-premise GPU instances. Such long-term GPU commitments not only ensure GPU availability at all times but also offer lower prices over extended periods compared to on-demand instances. For example, a three-year reserved instance for 8 \texttt{H100} GPUs (\texttt{p5.48xlarge}) on AWS costs \$$37.56$/hour, while the equivalent on-demand instance costs \$$98.32$/hour.
On-premise deployments can offer even greater cost savings: as shown in~\cite{on-prem-reserve-cost-comp}, they can reduce costs by up to 46.3\% over time compared to reserved cloud instances when accounting for lifetime return on investment.
As a result, this strategy has become the norm among today’s LLM service providers. For example, OpenAI has deployed tens of thousands of GPUs in its data centers~\cite{morikawa_2023, sherwood2024gpuhoarding}; similar deployments are also reported by
ByteDance~\cite{shilov2024bytedance} and xAI~\cite{xai-gpu-cluster}. 

\MyPara{Latency metric: Time-to-First-Token (TTFT).}
A critical performance metric for online serving is TTFT, defined as the latency from when a request initiates until the first output token is produced. When the client and model replica are in the same region, TTFT primarily consists of prefill latency and queuing delay. For example, when deploying on a L4 GPU, processing a $512$-token prompt with \btexttt{meta-llama/Llama-3.1-8B-Instruct} might incur around 300 ms of prefill latency before generating the first token. In a multi-region deployment, TTFT additionally includes cross-region network latency, which typically adds up to $200$ ms.

\begin{figure}[t]
  \centering
  \begin{subfigure}[t]{.63\linewidth}
  \includegraphics[width=\linewidth]{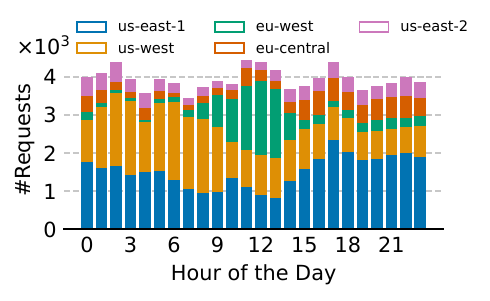}
  \caption{Regional load.}
  \label{regional-load-aggregated}
  \end{subfigure}
  ~
  \begin{subfigure}[t]{.36\linewidth}
  \includegraphics[width=\linewidth]{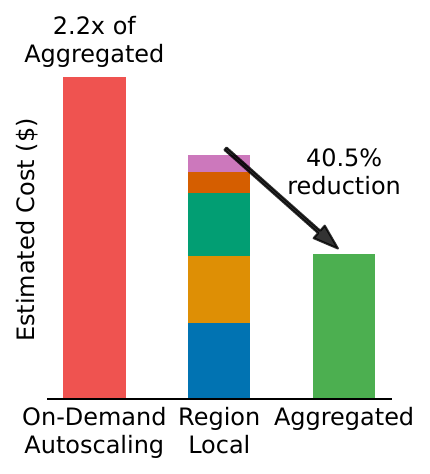}
  \caption{Cost comparison.}
  \label{cost-comparison-aggregation}
  \end{subfigure}
  \caption{(a) Aggregated load across five regions using a subset of the WildChat trace~\cite{zhao2024wildchat}. Before aggregation, per-region load variance ranges from $2.88\times$ to $32.64\times$; after aggregation, the variance is reduced to $1.29\times$. (b) Cost reduction achieved by provisioning based on the aggregated \emph{global} peak load, using reserved instances. We also present the cost for perfect on-demand autoscaling, assuming precise traffic prediction and no provisioning delay. In practical scenarios, the actual cost of on-demand autoscaling would be even higher.} 
  \label{fig:reduce-load-imbalance}
\end{figure}

\subsection{New Problem: Global Cost Reduction}
\label{ssec:cross-region-load-imbalance}

Multi-region LLM serving introduces new challenges for reducing serving costs. 
Our trace analysis reveals that multi-region LLM serving exhibits regional demand shifts over time, often following a diurnal pattern. Specifically, Figure~\ref{fig:diurnal-pattern} illustrates the load patterns of six countries from the WildChat trace~\cite{zhao2024wildchat}.  It shows clear diurnal trends---each region experiences peak inference traffic during specific hours, with lower load during the rest. These peak hours vary across regions due to time zone differences. 

The fluctuating load leads to time-varying GPU demand in each region, posing significant challenges for resource provisioning and cost reduction. Ideally, providers would provision just enough GPU resources in each region to meet the required GPU demand and adjust allocations dynamically as load varies. Recent studies attempt to approach this ideal by auto-scaling resources using on-demand or spot instances~\cite{mao2024skyserveservingaimodels, miao2024spotserve, patke2025hierarchical, zhang2019mark}. However, as discussed in Section \secref{ssec:muti-region-serve}, to avoid GPU shortages during demand spikes and to secure lower instance pricing, current multi-region deployments typically rely on static provisioning via long-term commitments. This makes them unable to scale flexibly with load fluctuations. As a result, to maintain service quality, providers must provision each region for its peak load. While this approach ensures low latency and high throughput, it inevitably leads to low overall GPU utilization and higher operational costs.

To illustrate this, Figure~\ref{regional-load-aggregated} shows the aggregated number of requests received across five regions at different hours of the day, using a subset of the WildChat trace~\cite{zhao2024wildchat}. It highlights that while regional loads fluctuate significantly, the aggregated global load remains relatively stable. As shown in Figure~\ref{cost-comparison-aggregation}, provisioning based on the peak aggregated global load can reduce costs by $40.5\%$ compared to provisioning each region for its own peak independently.
With on-premise instances, the cost can be reduced even further compared to reserved instances~\cite{on-prem-reserve-cost-comp}.
The figure also includes the cost of using on-demand instances. Due to their high pricing, even with perfect auto-scaling, where we assuming no provisioning delay and instances are always available, this approach incurs $2.2\times$ the cost compared to provisioning the aggregated global load using reserved instances.

\MyPara{Insight.} The cost reduction shown in Figure~\ref{cost-comparison-aggregation} suggests that we should provision instances based on \textit{global} demand and share these resources across regions. Users can be routed to access their geographically closest region, but when it becomes overloaded, traffic can be rerouted to remote regions to utilize otherwise unused capacity.

\subsection{Cross-Region Load Balancing}
\label{ssec:background-cross-region-load-balancing}

Achieving global cost reduction requires cross-region load balancing, a direction that remains largely unexplored in the context of LLM serving. Today's LLM load balancers are designed for single-region deployments and cannot be directly extended to multi-region settings. 
As shown in Figure~\ref{fig:overview}(b), a straightforward approach is to deploy existing load balancers in a specific zone as global coordinators. However, this creates centralized bottlenecks and a single point of failure, while also incurring non-trivial latency overhead due to cross-region communication for most requests.
\label{single-lb-large-cross-region-latency}
In typical local deployments, the TTFT latency for LLMs is mainly bottlenecked by prefill time and is usually several hundred milliseconds~\cite{stojkovic2024dynamollmdesigningllminference}, while cross-region latency can reach up to 200 milliseconds~\cite{mao2024skyserveservingaimodels}. This means that the straightforward approach may introduce additional latency nearly equivalent to the prefill time, significantly impacting responsiveness.
This motivates us to design a geo-distributed load balancer for LLM serving that can effectively reduce global serving costs without compromising TTFT. This involves addressing two key challenges:

\MyPara{KV Cache awareness.} 
LLM serving systems reuse KV Cache between requests that share the same prefix to reduce prefill time and saving compute resources.
To maximize the KV Cache hit rate, prior work has explored making load balancers prefix aware~\cite{sglang-router,srivatsa2024prebleefficientdistributedprompt,cao2025locality}. These approaches typically maintain a prefix tree for each replica, either precise, as in Preble~\cite{srivatsa2024prebleefficientdistributedprompt}, or approximate, as in SGLang Router~\cite{sglang-router} and DLPM~\cite{cao2025locality}.  This line of work has shown strong potential, and we observed up to a $35\%$ improvement in KV Cache hit rate in the best case.

However, extending them to multi-region deployments presents several challenges. Achieving KV Cache awareness requires maintaining a shared prefix tree, which is difficult to coordinate across regions. 
As previously discussed, simply maintaining a centralized global prefix tree will incur significant latency penalties due to frequent remote accesses.
Alternatively, maintaining fully distributed prefix trees in each region requires cross-region coordination on every request, which would become prohibitively expensive due to the high coordination overhead.

\begin{figure}[t]
  \centering
  \begin{subfigure}[t]{.48\linewidth}
    \centering
    \includegraphics[width=\linewidth]{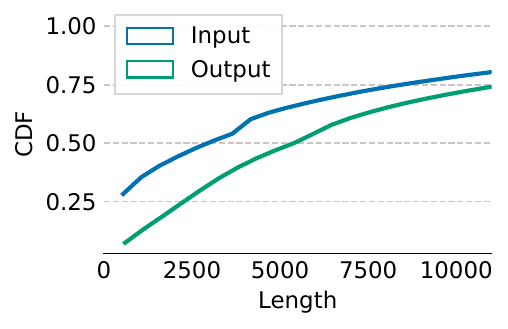}
    \caption{CDF for request length.}
    \label{fig:input-output-len-cdf}
  \end{subfigure}
  ~
  \begin{subfigure}[t]{.48\linewidth}
    \centering
    \includegraphics[width=\linewidth]{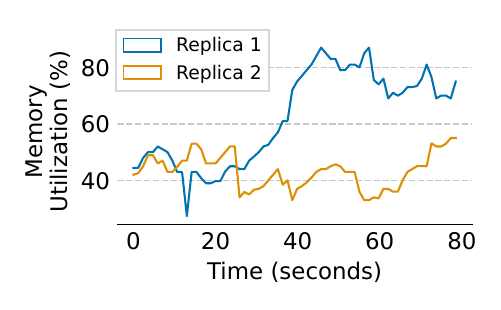}
    \caption{Load imbalance.}
    \label{memory-util-imbalance}
  \end{subfigure}
\caption{(a) The CDF of input length and output length in WildChat dataset. (b) We observe load imbalance across replicas when routing requests using the Round Robin algorithm. The y-axis shows the percentage of KV Cache memory utilization on each replica. The peak memory usage difference between replicas reaches $2.64\times$.}
  \label{fig:load-imbalance}
\end{figure}

\label{background-unknown-output-length}
\MyPara{LLM inference load unpredictability.}
The length of LLM outputs can vary widely, and in some cases be very long, as shown in real workloads (Figure~\ref{fig:input-output-len-cdf}) and prior studies~\cite{xiang2025servegenworkloadcharacterizationgeneration}. This variability, combined with the auto-regressive nature of LLMs, makes it difficult for load balancers to predict output lengths in advance and accurately estimate the GPU resources required for each request.  Moreover, each LLM request can demand substantial resources. It is not uncommon for a single request to consume several gigabytes of GPU memory, limiting each model replica to handling only tens of concurrent requests. Traditional load balancing polices, such as least-load-first or round-robin, blindly push requests to replicas without accounting for resource consumption, often resulting in long-running requests blocking all subsequent ones in the queue and causing severe load imbalance. These unique characteristics of the LLM workload make the \textit{penalty of misrouting} much more severe than in traditional CPU-based workloads. This motivates the need for a new system design that is robust to the dynamic and resource-intensive nature of LLM workloads.

\section{\sys Design}
\label{sec:skywalker-design}

\MyPara{Overview.}
In this work, we introduce \sys, a new design that enables efficient cross-region load balancing for LLM inference. With \sys, LLM service providers can reduce global costs by sharing a smaller pool of reserved or on-premise instances across regions without sacrificing performance. \sys deploys load balancers in multiple regions as the first point of contact for local requests, and introduces a cross-region traffic handler that coordinates traffic between regional load balancers to mitigate cross-region load imbalance (\secref{ssec:cost-efficiency-with-cross-region-traffic}). It preserves the benefits of prefix sharing by supporting prefix-aware routing in two ways (\secref{ssec:prefix-aware-routing}):
(1) a simple yet effective policy based on consistent hashing that requires minimal changes to existing load balancers; and
(2) prefix-aware routing using partial prefix tree snapshots maintained at each load balancer.
In addition, to address the challenge of LLM inference load unpredictability, \sys introduces a novel selective pushing mechanism that balances load based on pending requests at each replica (\secref{ssec:selective-pushing}).
We detail each of these design choices in the following.

\subsection{Cross-Region Traffic Handling}
\label{ssec:cost-efficiency-with-cross-region-traffic}

Since geographical regions could experience peak load at different times, we can exploit the traffic pattern by offloading workloads from high-demand regions to low-demand ones. 
This mitigates the load imbalance caused by regional demand shifts over time and reduces the total number of required GPU instances compared to region-local deployment.

There are multiple possible approaches to handling cross-region traffic. One approach is to deploy load balancers in a single zone that manage replicas across all regions. In this setup, requests from all regions are first sent to the central load balancer, which then routes them to replicas possibly in different regions. However, as discussed in \secref{single-lb-large-cross-region-latency}, this results in long round-trip times for users whose requests traverse multiple regions: the request incurs two cross-region RTTs, i.e., one to reach the load balancer and another to reach the assigned replica, leading to high cross-region latency. 

Another approach is to replicate the load balancer across multiple regions, allowing each to route traffic to all available replicas. In this deployment, clients send their requests to the nearest load balancer, which then decides which replica should handle the request, potentially in any region.
However, this approach requires non-trivial synchronization \textit{between load balancers} to make coordinated routing decisions. Without such coordination, multiple load balancers may independently select the same replica as hot spot, leading to degraded performance, particularly for affinity-aware load balancing strategies. 
Such coordination can incur non-trivial overhead for each load balancer, requiring $O(N_{\text{LB}} \times N_{\text{replica}})$ connections or probes. Whenever a new replica is launched, it must ensure that all load balancers are aware of its creation and updated status, adding complexity and latency to the system.

\MyPara{Our approach: two-layer cross-region routing.} 
Instead, we adopt a two-layer approach. The key idea is to coordinate cross-region traffic between load balancers, rather than directly between replicas.
Specifically, each load balancer either routes requests to local replicas or forwards them to other load balancers in remote regions, which then make the final placement decisions within their region.
This design combines the strengths of the two approaches discussed earlier. It avoids introducing significant routing latency by allowing clients to connect to their nearest load balancer, while still enabling cross-region load balancing through coordination among load balancers.
Comparing load balancers routing to all replicas distributed across regions, this approach significantly reduces the probing overhead required to assess replica load. In addition, managing connections between a small number of load balancers scales much better than maintaining connections to every replica, as the number of replicas typically far exceeds the number of load balancers.

\subsection{Multi-Region Prefix-Aware Routing}
\label{ssec:prefix-aware-routing}

The need for cross-region traffic routing presents new challenges for prefix-aware routing (\S\ref{ssec:background-cross-region-load-balancing}), especially with long cross-region latency. Achieving optimal prefix-aware routing requires a \textit{global} view of prefix states across all replicas. 
However, each load balancer only observes a \textit{subset} of requests, making it difficult to maintain a consistent global view without incurring significant coordination overhead on every request.
Thus, we ask the question: How can we effectively load balance across \textit{multiple regions} with prefix awareness? 
We answer this question by first understanding prefix sharing patterns in real workloads.

\begin{figure}[t]
  \centering
  \begin{subfigure}[t]{.56\linewidth}
  \includegraphics[width=\linewidth]{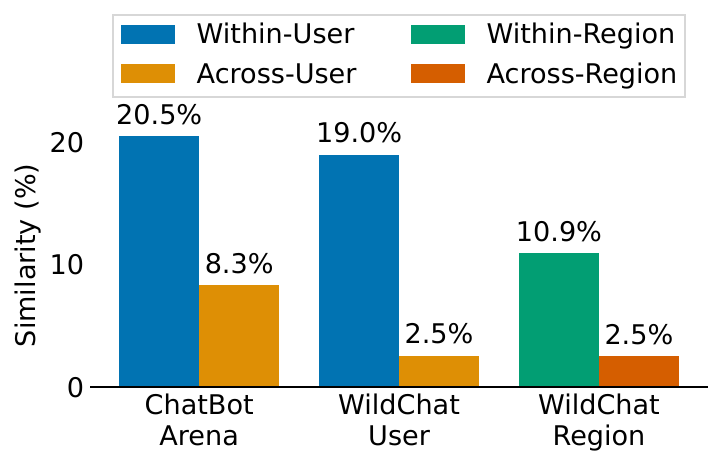}
  \caption{Prefix similarity.}
  \label{similarity-comparison-within-across}
  \end{subfigure}
  ~
  \begin{subfigure}[t]{.44\linewidth}
  \includegraphics[width=\linewidth]{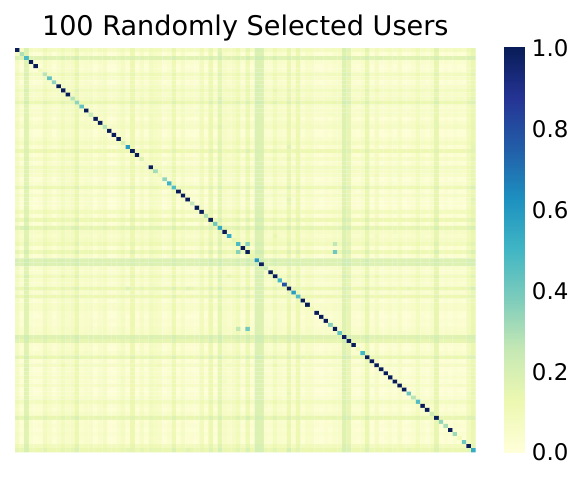}
  \caption{Similarity heatmap.}
  \label{similarity-heatmap}
  \end{subfigure}
  \caption{(a) Average prefix similarity within and across users and regions; (b) Heatmap of pairwise prefix similarity among 100 randomly sampled users, in WildChat~\cite{zhao2024wildchat} and ChatBot Arena~\cite{zheng2023judging} datasets. User information is retrieved directly from the metadata provided in each dataset (hashed IP in WildChat and judge in ChatBot Arena).}
  \label{fig:similarity-comparison}
  \vspace{-2mm}
\end{figure}

\label{prefix-similarity-analysis}
\MyPara{Prefix similarity analysis.} 
We analyze prefix similarity\footnote{We define the prefix similarity between two requests $a$ and $b$ as ${\text{len}(\text{common\_prefix}(a, b))}/{\min(\text{len}(a), \text{len}(b))}$. We use the minimum length in the denominator so that, for example, if $a$ is a prefix of $b$, the prefix similarity of $a$ and $b$ should be $1$.} using a subset of the WildChat~\cite{zhao2024wildchat} and ChatBot Arena~\cite{zheng2023judging} datasets. The goal is to quantify how much prefix reuse occurs in real-world workloads, which directly impacts the effectiveness of KV Cache reuse in LLM serving systems.
This metric measures the normalized length of the common prefix shared between two requests. We compute prefix similarity across all pairs of requests within the same user, and across different users.
The results are shown in Figure~\ref{similarity-comparison-within-across}. We observe that the average prefix similarity within the same user is significantly higher than that across different users ($2.47$-$7.60\times$ more). This pattern is also evident in the heatmap (Figure~\ref{similarity-heatmap}), which shows the similarity among 100 randomly selected users, further confirming that within-user requests are more likely to share context and thus benefit from prefix caching. However, there is still some degree of cross-user prefix similarity, and the relative ratio between within-user and cross-user prefix similarity is workload-dependent ($2.47\times$ for ChatBot Arena and $7.60\times$ for WildChat). This observation motivates us to present the following two solutions: \sys-CH and \sys. 
\sys-CH is simple and captures user-level prefix similarity. \sys captures both within-user and cross-user prefix similarity. We detailed the algorithm in Listing~\ref{alg:handle-request}.

\MyPara{\sys-CH.} 
\sys-CH uses consistent hashing~\cite{consistent-hashing} on user-provided keys (e.g., user ID, session ID) and routes a user request to a corresponding replica (Listing~\ref{alg:handle-request}, lines 23-26). \sys-CH is \textit{implicitly} prefix-aware: requests from the same user tend to share similar prefixes (e.g., context, chat history) and consistent hashing will map them to the same replica. 
\sys-CH adopts a ring hash~\cite{stoica2003chord} scheme, where each virtual node on the hash ring is assigned to a replica and each replica can have multiple virtual nodes, allowing balanced key distribution across replicas. We make two extensions to the traditional consistent hashing.
First, due to \sys's two-layer load balancing design, \sys-CH performs consistent hashing at both layers: the load balancer routes requests to other balancers based on consistent hashing, and each balancer applies consistent hashing to assign the request to one of its managed replicas as well.
Second, virtual nodes are skipped based on the availability of its associated replica (detailed in \secref{ssec:selective-pushing}, and Listing~\ref{alg:handle-request}, line 26). When that happens, the algorithm continues iterating over successive virtual nodes on the ring. \sys-CH requires minimal state maintained at load balancers, and can be easily incorporated into the existing software stack. 

Since \sys-CH focuses only on within-user prefix similarity, there are cases where \sys-CH falls short of being optimal. We discuss them in the following:
\label{ch-suboptimal}

\begin{itemize}[leftmargin=*, nosep, topsep=6pt]
    \item \textbf{Cross-User Prefix Sharing}: Requests from different users can share common prefixes (Figure~\ref{fig:similarity-comparison}), but \sys-CH can route these requests to two different replicas, missing the benefits of routing these requests to the same replica to maximize KV Cache hit rate.
    \item \textbf{Bursty Request}: Consistent hashing either hashes a given key to a single replica, or to a replica set. In the former, a request burst can overload the single replica. In the latter, consistent hashing misses the opportunities to leverage prefix-sharing for requests sent by the same user to different replicas in the replica set. 
    \item \textbf{Heterogeneous User Program}: If requests from a single user's program contain multiple patterns and lack consistent prefix structures, using the user ID (or session/program ID) as the hash key fails to exploit prefix reuse. Worse, it may route dissimilar requests to the same replica, increasing the risk of overloading that replica.
\end{itemize}

\begin{figure}[t]
  \centering    \includegraphics[width=.6\linewidth]{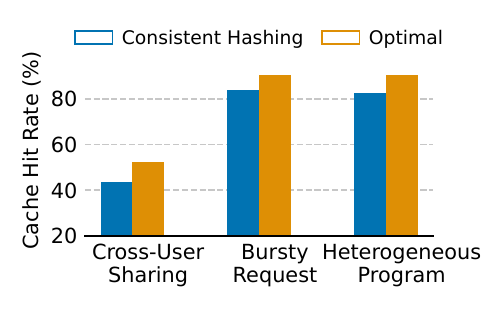}
  \caption{\textbf{KV Cache hit rate}, comparing consistent hashing with optimal solution with a global view.}
  \label{fig:prefix-policy-comparison}
\end{figure}

We illustrate these scenarios in Figure~\ref{fig:prefix-policy-comparison}, which shows the KV Cache hit rate under each setting, compared to an optimal solution with a global view. 
The lack of cross-user prefix sharing results in a $16.49\%$ prefix hit rate drop. We also study behaviors under bursty request patterns, where the variance in a single user's concurrent requests can reach $4\times$. In this case, CH leads to $7.07\%$ lower hit rate. A similar trend is observed when users submit heterogeneous programs, resulting in a $8.78\%$ gap. 
These limitations motivate us to develop \sys that is more general than \sys-CH by maintaining more states at the load balancer. 

\begin{algorithm}[t]\small
\caption{\sys load balancing logic}
\label{alg:handle-request}
\begin{algorithmic}[1]

\Procedure{MonitorAvailability}{}
  \While{true}
    \ForAll{$r \in \textit{LocalReplicas}$}
      \State $n_\text{pending} \gets \Call{ProbeReplica}{r}$
      \If{$n_\text{pending} > 0$}
        \State \Call{Remove}{LocalAvail, $r$}
      \Else
        \State \Call{Add}{LocalAvail, $r$}
      \EndIf
    \EndFor

    \ForAll{$lb \in \textit{RemoteLBs}$}
      \State $(n_\text{avail\_replica}, size_\text{q}) \gets \Call{ProbeLB}{lb}$
      \State \Comment{$\tau$: small buffer for newly arriving requests}
      \If{$n_\text{avail\_replica} = 0 \lor size_\text{q} > \tau$}
        \State \Call{Remove}{RemoteAvail, $lb$}
      \Else
        \State \Call{Add}{RemoteAvail, $lb$}
      \EndIf
    \EndFor
    \State \Call{Sleep}{ProbeInterval}
  \EndWhile
\EndProcedure

\Procedure{SelectCandidate}{Request, C}
  \If{UsePrefixTree}
    \State Text $\gets$ \Call{GetText}{Request}
    \State Trie $\gets \begin{cases}
               \text{ReplicaTrie} & \textbf{if } \text{C is replicas}\\
               \text{LBSnapSnotTrie} & \textbf{otherwise}
             \end{cases}$
    \State \Return \Call{MaxPrefixMatch}{Trie, Text, C}
  \Else
    \State HashRing $\gets \begin{cases}
               \text{ReplicaRing} & \textbf{if } \text{C is replicas}\\
               \text{LBRing} & \textbf{otherwise}
             \end{cases}$
    \State Key $\gets$ \Call{SessionId}{Request}
    \State HashValue $\gets$ \Call{Hash}{Key}
    \State \Return \Call{Next}{HashRing, HashValue, C}
  \EndIf
\EndProcedure

\Procedure{HandleRequest}{Request}
  \State $C \gets \begin{cases}
           \text{LocalAvail} & \textbf{if } LocalAvail \neq \emptyset\\
           \text{RemoteAvail} & \textbf{otherwise}
         \end{cases}$
  \State $t \gets \Call{SelectCandidate}{Request, C}$
  \State \Call{Route}{Request, t}
\EndProcedure

\end{algorithmic}
\end{algorithm}

\MyPara{\sys with regional snapshot.} \sys is \textit{explicitly} prefix-aware: in this design, each load balancer maintains prefix trees to keep an approximate view of prefix information on the load balancing targets. 
Between load balancers, the targets are remote load balancers, and between the load balancer and the replica, the targets are local replicas managed by that load balancer.

The prefix tree is a logical trie augmented with metadata to track active load balancing targets at each node. Each node stores a set of active targets associated with the prefix formed by the path from the root to that node. The tree is built incrementally from the requests the load balancer has served: when a new request is forwarded, a corresponding path is added to the trie, and the selected target is recorded at every node along that path. To bound memory usage, \sys enforces a configurable maximum tree size and evicts entries when the tree exceeds this limit, starting with the earliest inserted records.
\sys filter targets based on whether it is available to serve requests and pick the available target with the longest matching prefix (detailed in \secref{ssec:selective-pushing}, and Listing~\ref{alg:handle-request}, line 21). 
Specifically, for each trie traversal step, if there is no \textit{available} matching load-balancing target in the current node, the traversal terminates early. This is because the set of targets stored in any child node is always a \textit{subset} of its parent’s, implying that no available replicas can be found further down the path.

Each load balancer maintains two prefix trees, one for local replicas it manages and one for a partial view (snapshot) of other load balancers in other regions. The latter keeps track of all historical requests that the local region has sent to remote regions. Regional snapshots do not strictly record all prefixes reside in replicas of remote regions, which depends on requests that are sent to the load balancer of that region, either directly or from other load balancers. Instead, it is an approximation of prefixes \textit{that are possible to be utilized by local region forwarding to that remote region}, as we observe empirically that the local region is unlikely to share prefixes with requests that came from other regions: in Figure~\ref{similarity-comparison-within-across}, requests across regions only have $2.5\%$ prefix affinity.
With that, we observe \sys more closely approaches the performance of an optimal solution.

\subsection{Selective Pushing to Mitigate Load Imbalance}
\label{ssec:selective-pushing}

While leveraging prefix-affinity improves performance, it also leads to load imbalance as requests are preferentially routed to specific replicas. 
To address this, prefix-aware routing must be combined with effective load balancing strategies---for example, switching to a load-balancing policy when the prefix sharing ratio falls below a certain threshold. However, traditional load balancing strategies, such as blind pushing and selective pushing based on the maximum number of outstanding requests, do not work effectively for LLM workloads due to their load unpredictability (\S\ref{ssec:background-cross-region-load-balancing}). We begin by analyzing these two strategies and show how they lead to load imbalance when applied to LLM inference workloads. We then present our approach, selective pushing based on pending requests, as a solution to this problem.

\MyPara{Blind pushing.} 
One traditional load balancing strategy is to route each request to a replica \textit{immediately} upon arrival~\cite{der_Boor_2022,srivatsa2024prebleefficientdistributedprompt, sglang-router}, which we refer to as \textit{blind pushing}. Blind pushing performs well in CPU-based workloads or scenarios with uniform request processing times, where simple strategies like round-robin or least-load-first naturally result in a balanced load. However, the processing time of LLM varies from request to request, as it depends on the output length, which is difficult to predict due to the auto-regressive nature of decoding (\secref{background-unknown-output-length}).
Naively assuming each request is homogeneous can lead to a significant load differences across replicas. For example, we find two replicas under round robin can have memory usage difference up to $2.64\times$, as shown in Figure~\ref{memory-util-imbalance}. This issue is especially problematic when routing requests among multiple \emph{overloaded} replicas. Replica overload can occur in practice, as replicas are kept at high utilization to achieve better cost efficiency. During sudden surges in demand, request queues may accumulate, and a replica with a seemingly short queue may still incur long processing times if the requests in the queue takes long time to process. Blindly pushing requests to such overloaded replicas can lead to cases where requests waiting in one replica's queue while other replicas have idle compute capacity, wasting compute resources.

\MyPara{Selective pushing.} 
To address the unpredictability of LLM inference, we suggest selective pushing, a strategy where requests are temporarily queued at the load balancer and sent only to replicas that meet certain conditions. Specifically, the load balancer will only push requests to the replica that has capacity (decided by a threshold), and in the event that all replicas are full, queuing requests at the load balancer. 
This approach prevents overloading any single replica while maximizing overall utilization across all available replicas, ensuring that no request waits in one replica’s queue while others have idle compute capacity.
We explain two thresholds, outstanding requests and pending requests for selective pushing, and show that the latter is preferred in LLM serving. 

\MyPara{Selective pushing by limiting outstanding requests.} 
In this method, the load balancer selectively pushes to a replica only when the number of of outstanding requests for that replica is less than a \textit{fixed} threshold~\cite{rayServe2025, envoyLoadBalancers2025, kubernetesNetworking2025}. Each replica will not exceed its desired level of load and the rest will be queued in the load balancer. That way, when the request finishes on a replica and releases free capacity, it will inform the load balancer so that new requests are permitted to be served at this replica. 
However, selective pushing based on a fixed number of outstanding requests is ineffective for LLM workloads, since the number of requests a replica can serve depends on the total memory footprint of all outstanding requests, which is proportional to the total number of input and output tokens.
As the number of output tokens cannot be predicted in advance, the same inference engine can host a small number of large requests or a large number of small requests. We observed that for Llama 3.1 8B on a L4 GPU, the max number of outstanding requests can range from 20 to 50 for the same dataset. 
Therefore, statically setting the maximum threshold of outstanding requests delivers poor performance for LLM service (\secref{microbenchmark-sel-pushing}).

\MyPara{Selective pushing by checking pending requests.} 
We propose selective pushing by checking \textit{pending requests}. A \textit{pending request} is a request that has not been scheduled to the continuous batch yet, which indicates that the current batch is full and cannot admit more requests, as constrained by GPU memory. We use \textit{the existence of} pending requests in the replica to decide whether a replica is full or not. A background heartbeat probe is periodically sent to replicas to obtain their pending queue size (Listing~\ref{alg:handle-request}, line 3-8). If a replica has no pending request, it is ready to serve more requests. 

\MyPara{Selective pushing and cross-region routing.} Each load balancer tracks the number of replicas it manages with full continuous batches and periodically synchronizes this state with peer load balancers through heartbeat messages (Listing~\ref{alg:handle-request}, line 9-15). If a load balancer has at least one non-full replica and its request queue size does not exceed a small buffer (line 12), it is considered available to accept additional requests. Whether a peer load balanacer is available to serve requests is used to guide cross-region routing. When at least one local replica is not full, requests are always routed locally to maximize responsiveness. If all local replicas are full, the system considers remote regions and forwards requests only to regions with available replicas and short load balancer queue (Listing~\ref{alg:handle-request}, line 28). When multiple candidates are available, either among local replicas or remote load balancers, the system breaks ties using the consistent hashing key (for \sys-CH) or the prefix hit rate (for \sys) to select a candidate with more prefix sharing, as detailed in Listing~\ref{alg:handle-request}, line 17-26.

\section{\sys Implementation}

\begin{figure}[t]
  \centering
    \includegraphics[width=\linewidth]{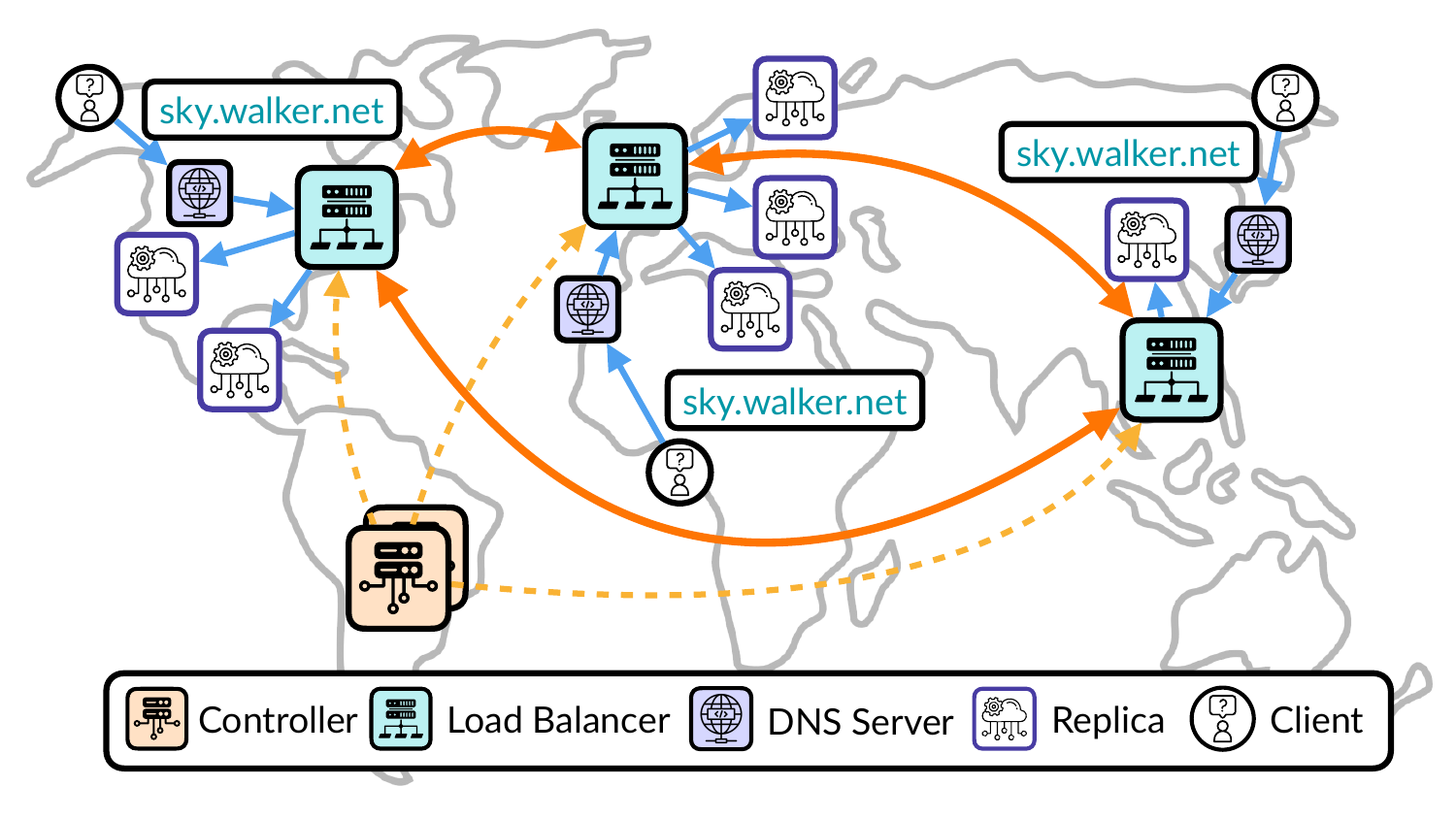}
  \caption{System architecture of \sys. Clients across different geographical regions issue requests to a unified domain name (\texttt{sky.walker.net} in this example). The domain will be resolved to the nearest available load balancer based on the client's IP. Each load balancer maintains connections to local replicas and other remote load balancers and performs request routing based on its load and shared prefix (\S\ref{ssec:prefix-aware-routing}).  Load balancers coordinate with each other to handle load redistribution. A centralized controller manages system updates, monitors health via periodic probes, and orchestrates failure recovery across all load balancers and replicas. Note that the placement of load balancers and replicas in the figure is illustrative; their geographic positions are abstracted for clarity and do not reflect exact deployment coordinates.}
  \label{fig:system-architecture}
\end{figure}

We implemented \sys (Figure~\ref{fig:system-architecture}), a prototype system that leverages geo-distributed load balancers to achieve both high throughput and low latency for online LLM serving across multiple geographical regions. \sys is built on top of SkyServe~\cite{mao2024skyserveservingaimodels}, an open-source multi-region serving framework for AI models, which supports both on-premise and cloud-based replicas. \sys extends SkyServe by adding geo-distributed load balancer with $\approx3000$ lines of Python code, and is compatible with any inference engine with OpenAI API, such as vLLM~\cite{kwon2023efficientmemorymanagementlarge} and SGLang~\cite{zheng2024sglang}.

\subsection{Load Balancer}
\MyPara{Load balancer distribution.} \sys launches load balancers in user-specified regions, with the number of instances configurable by the user. In cloud deployments, the regions are automatically inferred from the cloud region. \sys creates an AWS Route53~\cite{awsRoute53} DNS record for each load balancer, all associated with the same domain name for a unified endpoint. Under the hood, DNS resolution maps the domain name to the nearest load balancer based on the client’s source IP, thereby routing client requests to the nearest one.

\MyPara{Request life cycle.} A request first contacts the DNS server to resolve the IP address of the nearest load balancer. It is then sent to the load balancer and placed in a first-come, first-served (FCFS) request queue. When the request reaches the head of the queue, \textit{available} replicas in the local region are prioritized. If no local replica is available, the request is forwarded to an \textit{available} remote region with the highest prefix hit ratio. If the request is routed to a remote region, its snapshot is updated using the input prompt of this request.

\MyPara{Probe frequency.} We set the probing interval for selective pushing (\secref{ssec:selective-pushing}) to 100ms. This choice is motivated by the observation that each continuous batching step typically requires only tens of milliseconds to complete~\cite{llm-inference-benchmark}. A 100ms interval achieves a balance between responsiveness and stability: it is short enough to capture meaningful changes in system load, yet long enough to avoid excessive probing overhead.

\MyPara{Customized routing policy.} \sys supports customizable routing policies that ensure compliance with GDPR (\secref{gdpr}). For instance, \sys can restrict traffic offloading to occur only between GDPR-compliant regions, while regions outside the scope of GDPR may still offload traffic to GDPR-compliant regions when their replicas are underutilized.

\subsection{Serving System}
\MyPara{Service controller.} \sys employs a centralized controller to manage any updates to the system, such as adding or removing replicas and reconfiguring load balancers. It also performs periodic probes to monitor the status and health of all replicas and load balancers. The controller is fault-tolerant and can recover its state automatically in the event of a failure.

\MyPara{Failure recovery of load balancers.} \sys supports automatic recovery for load balancers. Upon detecting an unexpected failure via periodic health probes, the controller reassigns the replicas in the affected region to a geographically closest load balancer. That load balancer will then temporarily treat those replicas as local replicas. In parallel, a recovery process is initiated in the background. Once the failed load balancer is recovered, the associated replicas are transferred back to it. \sys can tolerate multiple concurrent load balancer failures. For higher fault tolerance, users can deploy additional load balancers in any region.

\section{Evaluation}

We evaluate \sys comprehensively across a variety of workloads and configurations to answer three questions:

\begin{itemize}[leftmargin=*, nosep, topsep=6pt]
    \item Can \sys maintain high throughput while preserving low latency in a geo-distributed setup? (\secref{macrobenchmark})
    \item What performance gains does selective pushing with pending requests (\secref{ssec:selective-pushing}) provide? (\secref{microbenchmark})
    \item What performance and cost benefits does \sys provides for regionally imbalanced workloads compared to standard region-local deployments? (\secref{microbenchmark}) 
\end{itemize}

\subsection{Macrobenchmarks}
\label{macrobenchmark}

We conducted end-to-end experiments using up to 12 replicas in a multi-region setup, where both replicas and clients are distributed across three geographical regions. We compare \sys with several production and research systems:

\begin{itemize}[leftmargin=*, nosep, topsep=6pt]
    \item \textbf{GKE Gateway~\cite{google2025gatewayapi}}: GKE Gateway is a network gateway service that connects multiple GKE~\cite{google-gke} clusters to provide a unified endpoint. Under the hood, each request is routed to and handled by one of the clusters. 
    \item \textbf{Round Robin (RR)}: A stateless load balancer that distributes incoming requests in a round-robin fashion.
    \item \textbf{Least Load (LL)}: A load balancer that tracks the number of outstanding requests per replica and routes each new request to the replica with the least load.
    \item \textbf{Consistent Hashing (CH)}: A ring hash~\cite{consistent-hashing,stoica2003chord} based consistent hashing algorithm, using the user's IP address and session ID as hash key.
    \item \textbf{SGLang Router~\cite{sglang-router} (SGL)}: A prefix-aware load balancer that routes requests based on a cache-aware routing algorithm tailored to LLM workloads.
    \item \textbf{\sys-CH}: \sys using a ring-hash based consistent hashing policy.
    \item \textbf{\sys}: \sys using the prefix tree policy.
\end{itemize}

We modify SkyServe~\cite{mao2024skyserveservingaimodels} to support our four global coordinator baselines: RR, LL, CH, and SGL. For those baselines, a single load balancer is deployed in the US. For both variants of \sys and GKE Gateway, a load balancer is deployed in each region.

\begin{figure*}[t]
\centering
\begin{subfigure}{.24\linewidth}
    \centering
    \includegraphics[width=\linewidth]{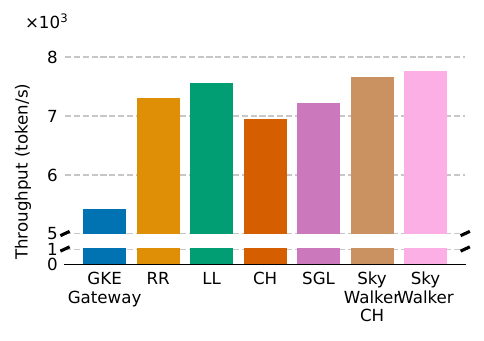}
    \caption{ChatBot Arena, Tpt.}
    \label{arena-tpt}
\end{subfigure}
~
\begin{subfigure}{.24\linewidth}
    \centering
    \includegraphics[width=\linewidth]{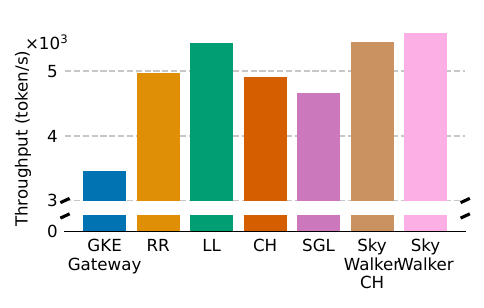}
    \caption{WildChat, Tpt.}
    \label{wildchat-tpt}
\end{subfigure}
~
\begin{subfigure}{.24\linewidth}
    \centering
    \includegraphics[width=\linewidth]{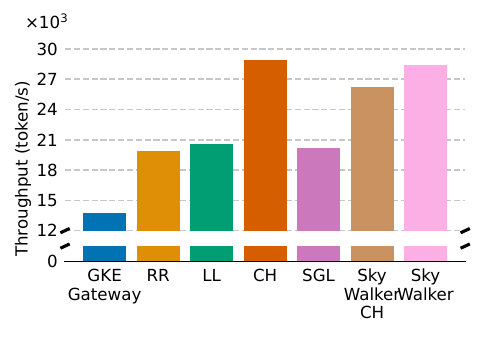}
    \caption{ToT, Tpt.}
    \label{tot-tpt}
\end{subfigure}
~
\begin{subfigure}{.24\linewidth}
    \centering
    \includegraphics[width=\linewidth]{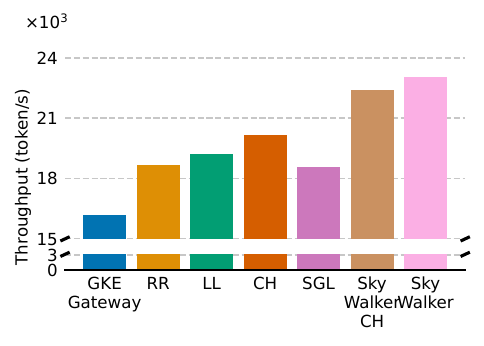}
    \caption{Mixed Tree, Tpt.}
    \label{tot-mixed-tpt}
\end{subfigure}
\\
\begin{subfigure}{.24\linewidth}
    \centering
    \includegraphics[width=\linewidth]{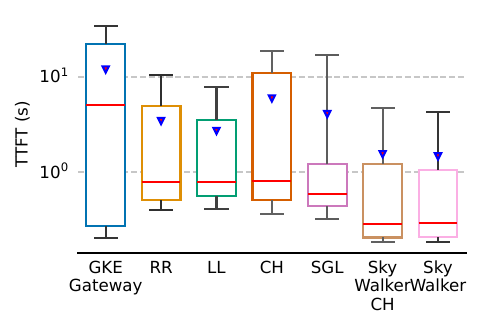}
    \caption{ChatBot Arena, TTFT.}
    \label{arena-ttft}
\end{subfigure}
~
\begin{subfigure}{.24\linewidth}
    \centering
    \includegraphics[width=\linewidth]{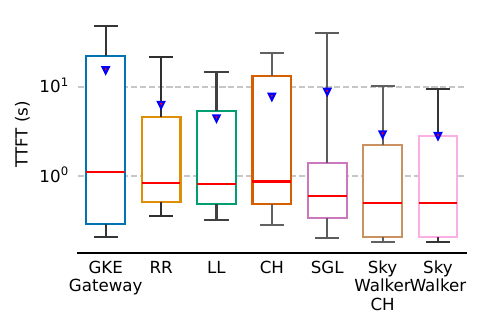}
    \caption{WildChat, TTFT.}
    \label{wildchat-ttft}
\end{subfigure}
~
\begin{subfigure}{.24\linewidth}
    \centering
    \includegraphics[width=\linewidth]{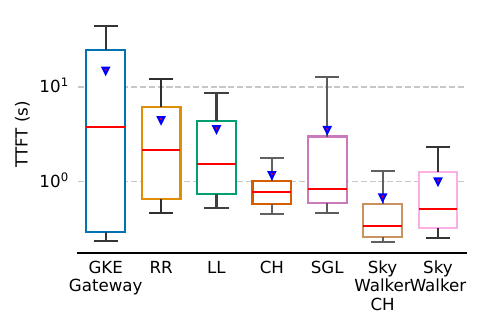}
    \caption{ToT, TTFT.}
    \label{tot-ttft}
\end{subfigure}
~
\begin{subfigure}{.24\linewidth}
    \centering
    \includegraphics[width=\linewidth]{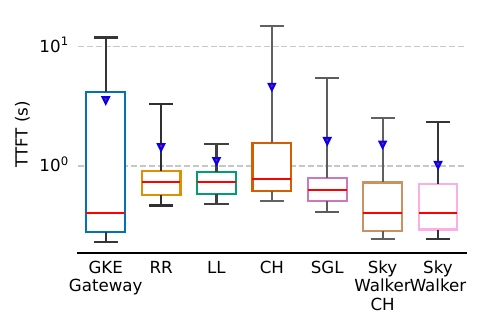}
    \caption{Mixed Tree, TTFT.}
    \label{tot-mixed-ttft}
\end{subfigure}
\\
\begin{subfigure}{.24\linewidth}
    \centering
    \includegraphics[width=\linewidth]{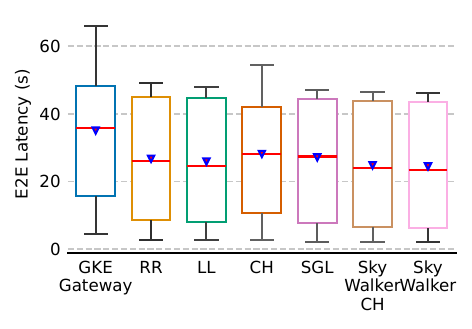}
    \caption{ChatBot Arena, E2E Latency.}
    \label{arena-e2e}
\end{subfigure}
~
\begin{subfigure}{.24\linewidth}
    \centering
    \includegraphics[width=\linewidth]{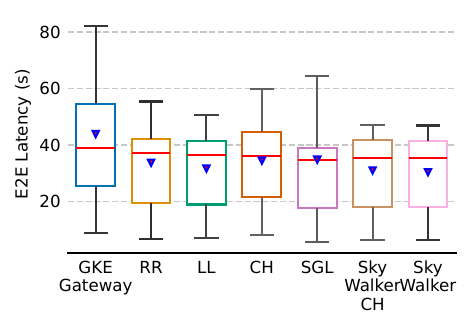}
    \caption{WildChat, E2E Latency.}
    \label{wildchat-e2e}
\end{subfigure}
~
\begin{subfigure}{.24\linewidth}
    \centering
    \includegraphics[width=\linewidth]{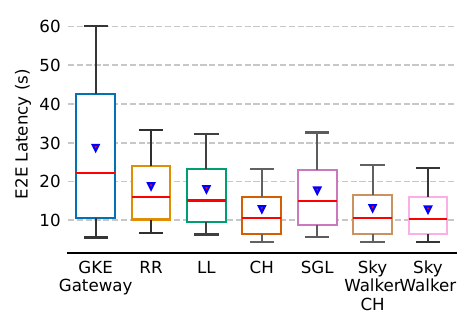}
    \caption{ToT, E2E Latency.}
    \label{tot-e2e}
\end{subfigure}
~
\begin{subfigure}{.24\linewidth}
    \centering
    \includegraphics[width=\linewidth]{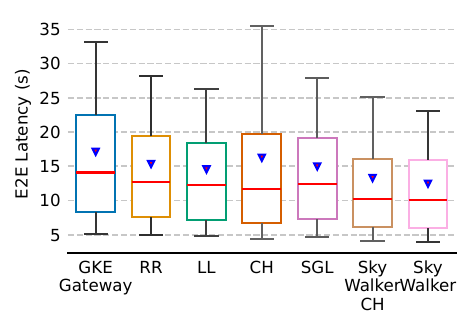}
    \caption{Mixed Tree, E2E Latency.}
    \label{tot-mixed-e2e}
\end{subfigure}
\caption{\textbf{Service Throughput, TTFT Latency, and End-to-End Latency.} We run \btexttt{meta-llama/Llama-3.1-8B-Instruct} on one L4 GPU with up to 12 replicas and report service throughput along with the distributions of TTFT and end-to-end latency. The TTFT latency plot is log-scaled. For the box plot, the red line marks the median, the box marks $25^{th}$ and $75^{th}$ percentiles, the whiskers show $10^{th}$ and $90^{th}$ percentiles, and the inverted triangle marks the mean.}
\label{fig:e2e-eval}
\end{figure*}

\MyPara{Experiment setup.} We conduct our evaluation primarily on
AWS~\cite{awsMainPage}, except for the GKE Gateway experiments which are
performed on GCP~\cite{gcpMainPage}. To ensure a fair comparison, each system uses the
same replica configuration. All replicas use one L4 GPU, hosting the \btexttt{meta-llama/Llama-3.1-8B-Instruct} model via SGLang~\cite{zheng2024sglang}. Replicas are distributed across three regions:
the United States, Europe, and Asia. We vary replica's
geographical allocation and client workload pattern to test a range of scenarios. For all experiments, we deploy clients in the US, Asia, and Europe to generate traffic, representing all end users in its respective region. Each client issues one program at a time. In practice, GPU utilization is kept high to maximize cost efficiency. In our evaluation, we maintain replicas at high utilization to reflect this real-world usage pattern. We use the following workloads:

\MyPara{Multi-turn conversation.} We evaluate all systems on several multi-turn conversation datasets and vary the client configuration to reflect different deployment scenarios:

\begin{itemize}[leftmargin=*, nosep, topsep=6pt]
    \item \textbf{ChatBot Arena~\cite{mao2024skyserveservingaimodels}}: A real-world multi-turn LLM conversation dataset collected using anonymized user IDs. For each region, we maintain the same number of clients, with 80 ongoing conversations per region. The real user ID in the dataset is used as its consistent hashing key.
    \item \textbf{WildChat~\cite{zhao2024wildchat}}: A large dataset of one million multi-turn conversations with demographic metadata such as state, country, and hashed IP address. We evaluate a configuration with different numbers of clients across regions: 40 in the US and 30 in both Europe and Asia. Each region issues requests only for conversations from its own geographical area, defined by the dataset's metadata. The hashed IP in the dataset is used as its consistent hashing key.
\end{itemize}

\MyPara{Tree of Thoughts.} We also evaluate on the Tree of Thoughts \cite{yao2023treethoughtsdeliberateproblem} benchmark using the Grade School Math dataset~\cite{cobbe2021gsm8k} from OpenAI. In this setting, the replica configuration is balanced, with 12 replicas evenly distributed across all regions (four per region). Tree of Thoughts exhibits high prefix reuse, as each question is solved via a tree structure where multiple nodes share prefixes from root to their least common ancestors. Nodes in the same tree can be executed concurrently. The tree has a depth of four, corresponding to a multi-step math reasoning task. The question ID in the dataset is used as the consistent hashing key. We evaluate two workload types:

\begin{itemize}[leftmargin=*, nosep, topsep=6pt]
    \item \textbf{Tree of Thoughts (ToT)}: Each tree uses a 2-branch structure (15 requests per tree). The US region runs 40 clients in parallel, while Europe and Asia run 20 clients concurrently.
    \item \textbf{Mixed Tree}: A more complex scenario where US runs 4-branch trees (85 requests per tree), with two clients sending such tree concurrently. The remaining regions continue to issue 2-branch trees, each with 20 clients in parallel. This setup reflects a mixed workload scenario where users generate heterogeneous traffic (e.g. setting different branch sizes for different accuracy requirements), more accurately representing real-world usage patterns. 
\end{itemize}

We report end-to-end service throughput, TTFT latency and end-to-end latency to evaluate system performance and responsiveness (Figure~\ref{fig:e2e-eval}).

\MyPara{Service throughput.} We show the service throughput of multi-turn conversation datasets (ChatBot Arena and WildChat) in Figrue~\ref{arena-tpt}, \ref{wildchat-tpt}. Both variants of \sys improve service throughput by $1.12$-$1.2\times$ compared to single load balancer solutions. Prefix-aware baselines such as SGL and CH rely on blind pushing, which leads to overloading some replicas while leaving others underutilized. In these baselines, replicas experience high variance in outstanding request counts, ranging from $2.33$-$5.08\times$ for SGL and $2.54$-$4.92\times$ for CH. Non-prefix-aware baselines perform worse in terms of prefix hit rate. RR achieves only $10.78$-$16.57\%$, while LL performs better with $28.29$-$31.13\%$, though still below \sys’s higher hit rate of $36.96$-$46.55\%$. Among the single load balancer baselines, LL achieves the best throughput, as load balancing plays a more dominant role when prefix similarity is relatively low. Nevertheless, it still falls short of \sys, reaching $97.38\%$ of \sys’s throughput.
Compared to GKE Gateway, \sys achieves a throughput improvement of $1.43$-$1.62\times$. This gain is primarily due to \sys’s LLM-specific design. While GKE Gateway offers robust, general-purpose multi-cluster load balancing, it lacks  prefix-aware routing for KV Cache optimization and the selective pushing mechanism that adapts to the dynamic nature of LLM workloads. The absence of these capabilities in a standard gateway solution results in lower cache hit rates ($18.08$-$24.30\%$) and less efficient GPU utilization, thereby constraining overall service throughput.

In the Tree of Thoughts workload (Figrue~\ref{tot-tpt}), when all trees are of uniform size, the CH baseline slightly outperforms \sys with a marginal throughput gain of $2\%$. CH also outperforms LL by $1.4\times$ in ToT, due to substantial prefix sharing. CH hashes requests from the same tree (i.e., the same question) to the same replica, enabling effective reuse of cached prefixes.
However, this advantage disappears under heterogeneous workloads (e.g., 2-branch vs. 4-branch trees, Figrue~\ref{tot-mixed-tpt}), and user-generated request bursts can saturate individual replicas. In such cases, the CH policy continues routing requests from the same user to the same replica, leading to significant overload with a variance in number of outstanding requests of $3.36\times$. SGL also suffers under this workload, showing high variance of $2.22\times$ as well. Non-cache-aware policies such as LL and RR experience low cache hit rates ($58.66$-$59.32\%$) compare to \sys's $89.56$-$90.01\%$, and consequently deliver suboptimal throughput.

Across all experiments, the prefix tree variant (\sys) consistently outperforms the CH variant (\sys-CH) by $1.34$-$8.21\%$. This is primarily because consistent hashing can occasionally assign users with bursty request patterns to the same replica (\secref{ch-suboptimal}), leading to load imbalance—since CH always routes requests to the same replica if it is available. In contrast, the prefix tree variant is more adaptive: when the prefix hit ratio is low (e.g., $<50\%$), it explores other underutilized replicas and distributes requests more evenly these replicas. This occasionally results in a slightly higher TTFT due to the added prefill time (e.g., in Figure~\ref{tot-ttft}), but it balances the load and delivers better overall throughput (Figure~\ref{tot-tpt}).

Compare to GKE Gateway, \sys offers key advantages through its KV Cache awareness and selective pushing mechanisms, which together contribute to $1.43$-$2.06\times$ higher service throughput. In contrast, a general-purpose gateway like GKE Gateway may incur longer prefill and queuing delays due to a lack of LLM-specific optimizations.

\MyPara{TTFT and end-to-end latency.} Regarding TTFT, the P50 and mean latency are primarily affected by cross-region latency and prefill latency. \sys achieves the lowest P50 and mean latency, ranging from $15.87\%$ to $57.63\%$ of the baseline values, across all evaluated systems (Figure~\ref{arena-ttft}-\ref{tot-mixed-ttft}). This improvement is attributed to its geo-distributed load balancers, which reduce cross-region latency, and its high prefix hit rates, which reduce prefill time. The P90 latency, largely determined by queuing delays, can reach several seconds. Even under this constraint, \sys maintains the lowest P90 TTFT ($10.08$-$23.38\%$ of baselines), owing to its selective pushing algorithm and reduced queuing delay.

For end-to-end latency (Figures~\ref{arena-e2e}-\ref{tot-mixed-e2e}), \sys consistently delivers the best performance, achieving $1.05$-$2.14\times$ improvements in P50 latency compared to baseline systems. This demonstrates that \sys effectively leverages KV Cache locality while maintaining balanced load.

\MyPara{Replica distribution.} We observe that \sys is robust to various replica distributions, including deployments with different numbers of model replicas and varying replica ratios across regions. In our end-to-end experiments, we evaluated different configurations, such as an unbalanced distribution (3 replicas in the US, 3 in Asia, and 2 in Europe) and a balanced distribution (4 replicas per region). \sys consistently achieves strong performance across all scenarios.

\subsection{Microbenchmark}
\label{microbenchmark}

\begin{figure}[t]
\centering
\begin{subfigure}{.32\linewidth}
    \centering
    \includegraphics[width=\linewidth]{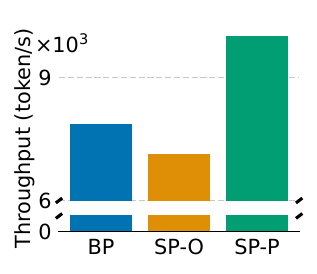}
    \caption{Throughput.}
    \label{ablation-pushing-tpt}
\end{subfigure}
~
\begin{subfigure}{.32\linewidth}
    \centering
    \includegraphics[width=\linewidth]{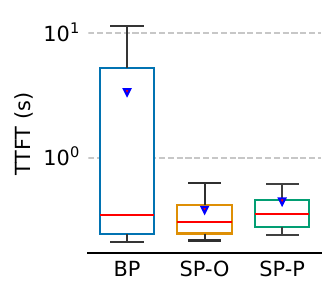}
    \caption{TTFT.}
    \label{ablation-pushing-ttft}
\end{subfigure}
~
\begin{subfigure}{.32\linewidth}
    \centering
    \includegraphics[width=\linewidth]{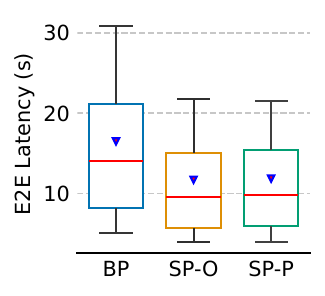}
    \caption{E2E latency.}
    \label{ablation-pushing-e2e}
\end{subfigure}
\caption{\textbf{Service Throughput and Latency}, comparing Blind Pushing (\btexttt{BP}) with two variants of Selective Pushing: fixed maximum outstanding requests per replica (\btexttt{SP-O}) and pending request (\btexttt{SP-P}). The TTFT latency plot is log-scaled. 
}
\label{fig:ablation-sel-pushing}
\end{figure}

\label{microbenchmark-sel-pushing}
\MyPara{Selective pushing by checking pending requests.}
We evaluate the effectiveness of the selective pushing mechanism (\secref{ssec:selective-pushing}) using one of our baseline systems, SGLang Router. We extend the original router to incorporate two variants of selective pushing: the standard one which is based on a fixed maximum outstanding requests per replica (\btexttt{SP-O}), and ours variant which is based on checking pending requests (\btexttt{SP-P}). We compare both against the original version that uses blind pushing (\btexttt{BP}). To isolate the effect of selective pushing, the experiment is conducted entirely within a single region, where all components (clients, replicas, and load balancer) are co-located. In this setup, TTFT is primarily influenced by prefill time and queuing delay.

The experiment uses 4 replicas and 30 clients within a single region, running the Tree of Thoughts (ToT) workload with a branching factor of 2. Results are shown in Figure~\ref{fig:ablation-sel-pushing}. \btexttt{SP-P} improves service throughput by $1.27\times$ (Figure~\ref{ablation-pushing-tpt}) and significantly reduces P90 TTFT by $18.47\times$ compared to \btexttt{BP} (Figure~\ref{ablation-pushing-ttft}). This demonstrates that \btexttt{SP-P} effectively minimizes queuing delay and improves prefill time by achieving a higher KV Cache hit rate: $89.86\%$ compared to \btexttt{BP}'s $68.89\%$. These gains translate into both lower latency (Figrue~\ref{ablation-pushing-ttft}, \ref{ablation-pushing-e2e}) and higher throughput. Compared to \btexttt{SP-O}, \btexttt{SP-P} achieves similar TTFT but improves throughput by $1.4\times$, highlighting that the adaptive nature of \btexttt{SP-P} leads to better replica utilization and overall performance under the same configuration.

\begin{figure}[t]
  \centering
    \includegraphics[width=.6\linewidth]{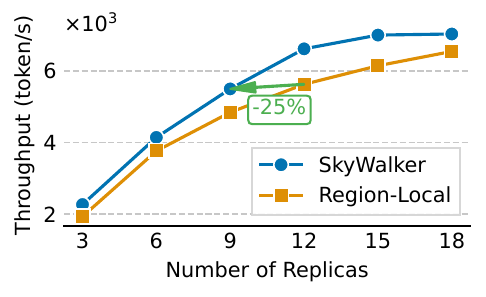}
  \caption{\textbf{Service Throughput,} comparing \sys and Region-Local deployments. We evaluate the performance using a regionally skewed workload, where the US region has 120 clients and both Asia and Europe have 40 clients. We vary the number of replicas to measure throughput gains from cross-region traffic offloading.}
  \label{fig:ablation-cross-region}
\end{figure}

\MyPara{Diurnal pattern.} We also evaluate \sys under regionally imbalanced workloads to assess its performance in handling traffic patterns with diurnal pattern. We compare it against a region-local deployment strategy where each region handles requests exclusively within its own local replicas, as is common among model providers (Figure~\ref{fig:overview}(a)). Specifically, we simulate a regionally skewed workload scenario representative of typical US working hours, where the US region uses 120 clients, while both Asia and Europe have 40 clients. We vary the total number of replicas deployed in both \sys and the region-local baseline, with replicas are evenly distributed across the three regions.

The throughput results for both systems are shown in Figure~\ref{fig:ablation-cross-region}. With an equal number of replicas, \sys consistently outperforms the region-local system by between $1.07\times$ and $1.18\times$, demonstrating the effectiveness of cross-region traffic handling to offload traffic onto regions with less load. Moreover, we observe that \sys achieves comparable throughput with only 9 replicas as the region-local deployment achieves with 12 replicas, translating into a cost reduction of $25\%$ while maintaining the same level of throughput. 

\section{Related Work}
\label{related-work}

\MyPara{Load balancing for CPU workloads.} Efficiently managing CPU resources for latency-sensitive applications is a well-studied problem in CPU workloads. Prior work proposes load balancing policies that make task distribution decisions at microsecond-scale latencies. McClure et al.~\cite{mcclure2022efficient} classify these techniques into two categories: \textit{work stealing}~\cite{nookala2024x, karsten2020user, fried2020caladan, ousterhout2019shenango, li2016workstealing, bugnion2017zygos,dinan2009scalable} and \textit{work shedding}~\cite{nandagopal2010sender}. In work stealing, idle CPU cores actively pull jobs from overloaded cores. In contrast, work shedding involves overloaded cores pushing excess jobs to other cores. Empirical studies show that work stealing generally outperforms work shedding in terms of both latency and CPU utilization.

\MyPara{Production systems.} Amazon Bedrock~\cite{aws2025bedrock} is a fully managed LLM inference service that supports cross-region inference to handle traffic spikes. However, its offloading is limited to within the same continent, missing the opportunity to aggregate diurnal patterns. Additionally, Bedrock is a hosted solution operating at AWS scale, whereas \sys is a self-hosted serving system designed for broader accessibility. GCP Gateway~\cite{google2025gatewayapi} provides a unified endpoint for global deployment by routing requests across multiple GKE~\cite{google-gke} clusters. However, this solution is not tailored for LLM workloads. Neither Bedrock nor GCP Gateway incorporates prefix awareness, thereby failing to reuse KV Cache and reduce compute overhead. They also lack selective pushing based on pending requests, making them more susceptible to replica overload.

\MyPara{Prefix-aware load balancing.} Prior work has explored leveraging KV Cache reuse to improve the efficiency of LLM request routing. Preble~\cite{srivatsa2024prebleefficientdistributedprompt} achieves low latency and high throughput by maximizing prefix cache hit rates, but it relies on a centralized global scheduler, limiting its applicability to a single-region setting. Similarly, SGLang Router~\cite{sglang-router} maintains a global prefix tree across all replicas, incorporating more fine-grained load balancing policies. DLPM~\cite{cao2025locality} introduces a scheduling algorithm that improves upon Preble in both latency and throughput while also offering fairness guarantees to clients. While these centralized approaches deliver high performance, their reliance on a single scheduler makes them unsuitable for cross-region, production-grade deployments due to high inter-region communication latency and the inherent risk of a single point of failure.

\MyPara{Improving GPU utilization through job colocation.} Many techniques has been proposed to improve GPU utilization by sharing resources either spatially or temporally~\cite{oliaro2024flexllm, dhakal2020gslice, zhao2023muxflow, xiao2020antman, xiao2018gandiva, mobin2023colti, choi2021multi, yu2025prism}. These include strategies such as colocating training and serving jobs or enabling multi-model serving on shared GPUs. However, due to the strict service-level objectives (SLOs) associated with serving tasks, job interference remains a concern that can degrade performance. Moreover, in many real-world settings, users may only run serving workloads, limiting the opportunities for job colocation.

\section{Discussion and Future Work}

\label{gdpr}
\MyPara{GDPR and regulatory constraints.} The General Data Protection Regulation (GDPR) is an EU law that governs the collection, processing, and transfer of personal data, granting individuals strong rights over their information. It prohibits transferring data outside EU regions without adequate protections. In this context, while global traffic aggregation is not feasible, aggregating traffic within EU regions or offloading non-EU traffic to EU regions can still yield substantial cost savings.

\MyPara{Security risk for prefix sharing.} Recent work demonstrates that optimizations in LLM serving can expose timing and size-based side channels exploitable by adversaries to infer sensitive user information~\cite{cryptoeprint:2025/167}. Similarly, KV cache sharing in multi-tenant serving environments enables cross-user leakage, where an attacker can reconstruct private prompts by exploiting cache reuse~\cite{DBLP:conf/ndss/WuZZWNWZ25}. While prefix sharing may introduce potential side-channel vulnerabilities, the problem suggested by these findings is orthogonal to the design of \sys and pertains to all systems employing KV cache sharing, including single region or single inference engine deployments. Ongoing research continues to explore such risks and develop effective solutions.

\MyPara{Support for heterogeneous accelerators.} 
Load balancing is more challenging in heterogeneous environments. 
While \sys currently focuses on homogeneous replicas, it can be extended to support heterogeneous accelerators, such as different GPU types or other hardware like TPUs~\cite{jouppi2023tpuv4opticallyreconfigurable} and AWS Inferentia~\cite{aws-inferentia}. Notably, the selective pushing by checking pending requests mechanism in \sys is inherently hardware-agnostic: it identifies overloaded replicas without relying on hardware-specific features, making it naturally compatible with heterogeneous settings. However, the prefix-aware routing and overall load balancing policies remain as an open question. 

\MyPara{More advanced policies.} Request characteristics, such as prompt length, can influence ideal routing strategies. For instance, shorter prompts incur lower prefill costs, making it more advantageous to route them to replicas with slightly lower load instead of prioritize prefix reuse. \sys can be extended to incorporate request-characteristic aware routing strategies,  dynamically adapting its decision-making process based on each request. 

\section{Conclusion}

Shifting regional diurnal patterns make cost-efficient LLM deployment in multi-region setups challenging. To address this, we propose cross-region traffic handling and present \sys, a locality-aware cross-region load balancer designed to overcome those limitations. By enabling cross-region coordination through geo-distributed load balancers, \sys improves GPU utilization and reduces serving costs, all while maintaining low latency and high throughput. To achieve this, \sys leverages prefix-aware routing to maximize cache locality and selective pushing to adaptively avoid overloaded replicas. Together, these techniques provide robust load balancing under the inherently bursty and unpredictable nature of LLM workloads. Through extensive evaluation across real-world and synthetic scenarios, we show that \sys consistently outperforms existing production and research systems, achieving $1.12$-$2.06\times$ higher throughput and $1.74$-$6.30\times$ lower latency, and $25\%$ cost savings compare to other systems.

\section*{Acknowledgement}

We are grateful to our shepherd, Abhishek Chandra, and the anonymous EuroSys reviewers for their constructive feedback and thoughtful suggestions. This work is in part supported by gifts from Accenture, AMD, Anyscale, Broadcom, Cisco, Google, IBM, Intel, Intesa Sanpaolo, Lambda, Lightspeed, Mibura, Microsoft, NVIDIA, Samsung SDS, and SAP.

\bibliographystyle{ACM-Reference-Format}
\bibliography{paper}

\end{document}